\documentclass[twocolumn,nofootinbib,superscriptaddress,amsmath,amssymb,aps,prd,preprintnumbers,floatfix]{revtex4-2}
\pdfoutput=1
\usepackage{amsbsy,esint,mathrsfs,commath,amsfonts,mathtools,amsthm}
\usepackage{graphicx}
\usepackage{bm}
\usepackage[dvipsnames]{xcolor}
\usepackage[utf8]{inputenc}
\usepackage{enumitem}
\usepackage{tikz}
\usepackage{tikz-3dplot}
\usetikzlibrary{arrows.meta, automata, matrix}
\usetikzlibrary{shapes,arrows,positioning}
\usetikzlibrary{calc}
\usetikzlibrary{decorations.pathmorphing}
\usetikzlibrary{shapes.geometric}
\usetikzlibrary{arrows,decorations.markings}
\usetikzlibrary{patterns}
\usepackage{hyperref}
\usepackage[hyphenbreaks]{breakurl}
\hypersetup{colorlinks=true,allcolors=Blue}
\newcommand{\real}{\mathbb{R}}
\newcommand{\cmplx}{\mathbb{C}}
\newcommand{\uene}{\text{U}(N)}
\newcommand{\sudos}{\text{SU}(2)}
\newcommand{\cierrex}{\vec{\mathcal{X}}}
\newcommand{\cierrey}{\vec{\mathcal{Y}}}

\newcommand{\tinypois}[2]{\{#1,#2\}}
\newcommand{\pois}[2]{\left\{#1,#2\right\}}
\newcommand{\flujopois}[3][1]{
	\if 1#1
	\text{e}^{#2\pois{#3}{\bullet}}
	\else
	\text{exp}\left(#2\pois{#3}{\bullet}\right)
	\fi
}
\newcommand{\lrangles}[1]{\langle #1 \rangle}
\newcommand{\bra}[1]{\langle #1 |}
\newcommand{\ket}[1]{| #1\rangle }
\newcommand{\braket}[2]{\lrangles{#1 | #2}}

\newcommand{\vespe}[2]{\bra{#2} #1 \ket{#2}}
\newcommand{\dket}[1]{\left|#1\right]}

\newcommand{\dbraket}[2]{\left[#1 | #2\right\rangle}

\begin{document}
	\title{The two-vertex model of loop quantum gravity: anisotropic reduced sectors}
	\author{Iñaki Garay} 
	\affiliation{Department of Physics and EHU Quantum Center,
University of the Basque Country UPV/EHU, Barrio Sarriena s/n, 48940, Leioa, Spain}  
	\author{Luis J. Garay} 
	\affiliation{Departamento de F\'{\i}sica Te\'orica e IPARCOS, Universidad Complutense de Madrid, 28040 Madrid, España}
	\author{Diego H. Gugliotta}
	\affiliation{Departamento de F\'{\i}sica Te\'orica e IPARCOS, Universidad Complutense de Madrid, 28040 Madrid, España}
    \preprint{IPARCOS-UCM-25-007}
	\begin{abstract}
        The so-called two-vertex model of loop quantum gravity  has been analytically studied in the past within a $\uene $ symmetry-reduced sector leading to a cosmological interpretation.
      In this work we study the simplest non-trivial two-vertex model (with four edges, i.e., $N=4$), using the spinorial formalism and twisted geometries to isolate the degrees of freedom and derive a canonical parametrization. We identify eight geometric parameters describing the polyhedral configurations and four twist angles characterizing the system's dynamics. Going beyond the U($N$) symmetry-reduced sector which can be interpreted as homogeneous and isotropic, we find three additional stable symmetry-reduced sectors: the privileged-direction sector, the bi-twist sector, and the inhomogeneous bi-twist sector. Each sector introduces degrees of anisotropy or inhomogeneity and expand the potential cosmological interpretations of the two-vertex model.
	\end{abstract}

	\date{January 24, 2025}
	\maketitle

	\tableofcontents
\section{Introduction}
The quantum structure of   spacetime is described by loop quantum gravity (LQG)~\cite{Thiemann_2007,Rovelli_2004,statusreport,Ashtekar_LQGreview} in terms of a Hilbert space formed by spin-networks, i.e.,  wave functions defined over graphs with spins attached to the edges and intertwiners to the vertices.
On the other hand, loop quantum cosmology (LQC) follows the same quantization procedure of LQG as a proposal for the quantization of the homogeneous and isotropic sector of general relativity~\cite{BojowaldLQC,Ashtekar_LQCstatus,agulloLoopQuantumCosmology2016}.

The implementation of the dynamics, the semiclassical limit of LQG, and the relation between LQG and LQC are still among the main open problems in the field. 
In this sense, the study of simple models constructed as truncations---that arise from considering only the spin networks associated with a given graph---of the theory has proven to be extremely useful~\cite{RoveliSpeziale,Rovelli_Vidotto_SteppingOutHomogeneity2008, inakiDynamics4,inakiReturn,BorjaGarayVidotto_learning,BenitoLivine:classicalsetting,alvaro2024}.
   
The LQG classical phase space associated with a given graph can be described using the spinorial formalism~\cite{LivineTambornino,LivineTambornino_QuantAmbig,SDT_spinortwistors}, which provides a parametrization in terms of spinors. It is a  useful mathematical tool for the study of LQG and, more concretely, suitable to study reduced models and implement the dynamics on them. 
Moreover, the spinorial formalism provides a parametrization for the twisted geometries formalism~\cite{FreidelSpeziale:geo, FreidelSpeziale:twist, RoveliSpeziale}, in which the LQG phase space is described in terms of a discrete geometry. 
Specifically, each spinorial configuration defines a set of polyhedra dual to the graph, and the faces of the polyhedra connected by a given link of the graph have same area, although they do not necessarily fit together, hence the name of twisted geometries.

The simplest non-trivial graphs to be considered are those consisting of two vertices linked by $N$ edges, that, within the twisted geometries formalism, represent a pair of  polyhedra with $N$ faces.
These models were studied in the past employing the spinorial formalism with a suitable dynamics, giving rise to the so-called two-vertex model ~\cite{inakiReturn,BorjaGarayVidotto_learning}. 
It was possible  to identify  a symmetry reduced sector, the $\uene$ sector, which is stable under  evolution and fulfills the properties that are naturally interpreted as homogeneity and isotropy on the graph.
Moreover, the $\uene$-reduced dynamics has been shown to correspond to the effective LQC dynamics ~\cite{BenitoLivine:classicalsetting,alvaro2024}. 
Based on this cosmological behavior of the $\uene$ sector, it is sensible to study further relations of the model with cosmological scenarios not only in the symmetry-reduced case but in the general case as well. 
This possibility was reinforced by a numerical analysis of the general case~\cite{ArangurenGarayLivine}, but an analytical study of the model outside the $\uene $ sector is still missing. 

In this work, we will focus on the two-vertex model with four edges---which is the simplest case considering non degenerate polyhedra---aiming to get some insight on its behavior outside the homogeneous and isotropic regime.
For this purpose, we will work within the spinorial formalism.
We will solve the constraints on the spinors and characterize the relevant (non-gauge) degrees of freedom of the model, relating them to the description in terms of twisted geometries. 
Then, we will use the results to gain new intuition on the $\uene $ reduction, and to explore the possibility of less constrained stable reduced sectors in which to study the model analytically.
    
The paper is structured as follows. In Section~\ref{sec:dosver} we present the description of the two-vertex model, using the spinorial  and the twisted geometries  formalisms. 
In Section~\ref{sec:cuatroar} we study the two-vertex, four-edge  model within the spinorial formalism, completely solving the constraints on the spinors and hence providing a canonical parametrization in terms of the relevant degrees of freedom. 
Additionally, from the framework of the twisted geometries,  the parameters are characterized in relation to recognizable geometric quantities. 
The $\uene $ reduction is analyzed  in Section~\ref{subsec:uene}  within the presented parametrization, recovering previous results in the literature ~\cite{BorjaGarayVidotto_learning, inakiDynamics4,inakiReturn,BenitoLivine:classicalsetting,alvaro2024}. 
Finally, Section~\ref{sec:sectores} presents three symmetry-reduced sectors found to be stable under evolution which contain the $\uene $ sector as a subsector. 
Their additional degrees of freedom (compared to the $\uene $ sector) admit a natural interpretation as degrees of anisotropy or inhomogeneity. We briefly summarize and conclude in Section~\ref{sec:Conclusion}.
    
\section{Two-vertex model}\label{sec:dosver}
In this work we will consider the so-called two-vertex model \cite{BorjaGarayVidotto_learning,inakiReturn,inakiDynamics4,ArangurenGarayLivine,BenitoLivine:classicalsetting,alvaro2024}, that is, a truncated model within LQG represented by graphs made of two vertices linked by $N$ edges.
In this section, we briefly describe the phase space of the model following the spinorial formalism \cite{LivineTambornino,inakiReturn}, together with the interpretation given in terms of the twisted geometries \cite{FreidelSpeziale:geo,FreidelSpeziale:twist}. 
We will also present the Hamiltonian proposed in \cite{inakiDynamics4,inakiReturn}, and further analyzed in~\cite{BenitoLivine:classicalsetting,ArangurenGarayLivine,alvaro2024}.

\subsection{Spinorial phase space}\label{subsec:efespin}
    
\begin{figure}
    \begin{tikzpicture}[node distance=3cm]
    \node[circle,draw](a){$\alpha$};
    \node[right=of a](c){{$\begin{aligned}
        \scriptstyle\bullet\\ \scriptstyle\bullet\\ \scriptstyle\bullet
    \end{aligned}$}};
    \node[circle,draw](b)[right=of c]{$\beta$};
    \path[draw] (a) ..controls ++(75:1) and ++(175:3).. 
    node[pos=0.4
,above=0.05cm]{\hspace{-11pt}$\ket{z_1}$}
    (3,2.5)
    ..controls ++(175:-3) and ++(85:3)..
    node[pos=0.6,above=0.05cm]{\hspace{10pt}$\ket{w_1}$}
    (b);
    \path[draw] (a) ..controls ++(40:1) and ++(180:1.5).. 
    node[ pos=0.35,above=0.10cm]{$\ket{z_2}$}
    (3.2,1.7)
    ..controls ++(180:-1.5) and ++(110:2)..
    node[near end,above=0.15cm]{$\ket{w_2}$}
    (b);
    \path[draw] (a) ..controls ++(20:2) and ++(180:1).. 
    node[near start,above=0.05cm]{$\hspace{12pt}\ket{z_3}$}
    (4,0.8)
    ..controls ++(180:-1) and ++(150:1)..
    node[pos=0.58,above=0.05cm]{$\ket{w_3}$}
    (b);
    \path[draw] (a) ..controls ++(-95:1) and ++(-175:3).. 
    node[pos=0.65,below=0.05cm]{\hspace{-8pt}$\ket{z_N}$}
    (2.5,-2.5)
    ..controls ++(-175:-3) and ++(-85:3)..
    node[pos=0.6,below=0.05cm]{\hspace{15pt}$\ket{w_N}$}
    (b);
    \path[draw] (a) ..controls ++(-50:1) and ++(-180:1).. 
    node[ midway,below=0.25cm]{$\ket{z_{N-1}}$}
    (2,-1.5)
    ..controls ++(-180:-1) and ++(-110:2)..
    node[near end,below=0.15cm]{$\ket{w_{N-1}}$}
    (b);
    \path[draw] (a) ..controls ++(-20:0.5) and ++(-190:1).. 
    node[near end,below=0.05cm]{$\hspace{4pt}\ket{z_{N-2}}$}
    (2,-0.6)
    ..controls ++(-190:-1) and ++(-150:2)..
    node[near end,below=0.05cm]{$\ket{w_{N-2}}$}
    (b);
\end{tikzpicture}
    \caption{Two-vertex model, with vertices $\alpha$ and $\beta$ linked by $N$ edges. 
        Each edge is associated with two spinors $\ket{z_i}$ and $\ket{w_i}$, one for each vertex.
    }
    \label{grafo}
\end{figure}
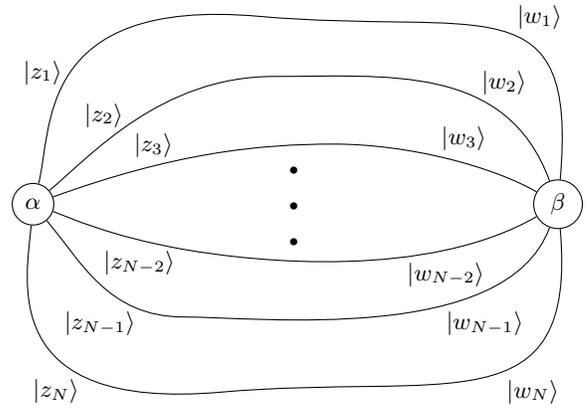

Consider a graph consisting of two vertices, $\alpha $ and $\beta $, linked by $N$ edges labeled by $i=1,2\dots N$ (see Fig.~\ref{grafo}). 
Within the spinorial formalism~\cite{LivineTambornino,inakiReturn} a spinor is attached to each edge and vertex of the graph. More explicitly, we attach a spinor $\ket{z_i}=\begin{pmatrix}
			z_i^0 & z_i^1
		\end{pmatrix}^\text{t}\in\cmplx^2
$ 
(the superscript t meaning transposition) to the $i$-th edge connected to the vertex $\alpha$.
For convenience, we also introduce dual spinors
$\dket{z_i}:=\begin{pmatrix}
			-\overline{z}_i^1& \overline{z}_i^0
		\end{pmatrix}^\text{t}
$ 
and  vectors $\vec{X}_i\in\real^3$ with components $X_i^a$ given by
\begin{equation}\label{vectores}
    X^a_i=\frac{1}{2}\vespe{\sigma^a}{z_i},
\end{equation}
where $\sigma^a$  are the usual Pauli matrices, normalized to $(\sigma^a)^2=\mathbb{I}$ for each $a=1,2,3$.
The norm of these vectors is
\begin{equation}
    X_i:=|\vec{X}_i|=\frac{1}{2}\braket{z_i}{z_i}.
\end{equation}
We make analogous definitions for the vertex $\beta $, where spinors are denoted by $\ket{w_i}$ and the corresponding vectors by $\vec{Y}_i$.

We endow these spinors with the symplectic structure defined by the Poisson brackets
\begin{equation}
    \pois{z_i^{A}}{\overline{z}_j^{B}}= \pois{w_i^{A}}{\overline{w}_j^{B}}=-i\delta_{ij}\delta^{AB},
\end{equation} 
with $A,B=0,1$, and the remaining Poisson brackets are zero.

Now we impose the $N$ matching constraints
\begin{equation}
    \mathcal{C}_i:=\braket{z_i}{z_i}-\braket{w_i}{w_i}=0,
\end{equation}
that, upon quantization, will ensure that the spin associated with each edge is vertex-independent \cite{LivineTambornino}.
It is straightforward to check that each of these constraints generates an action of the group $\text{U}(1)$ on the spinors at the corresponding edge:
\begin{equation}\label{gaugedeenlace}
    \flujopois{\theta}{\mathcal{C}_i}:(\ket{z_i},\ket{w_i})\mapsto (e^{i\theta}\ket{z_i},e^{-i\theta}\ket{w_i}),
\end{equation} 
while it commutes with the remaining spinors.
Note that the vectors $\vec{X}_i$ and $\vec{Y}_i$ are invariant under these transformations.
Additionally, we impose the six closure constraints (three for each vertex) 
\begin{equation}
    \vec{\mathcal{X}}:=\sum_{i=1}^N\vec{X}_i=0,\qquad \vec{\mathcal{Y}}:=\sum_{i=1}^N\vec{Y}_i=0.
\end{equation}
As will be discussed in Section \ref{subsec:geotuis}, these allow one to associate each vertex with a convex polyhedron whose faces correspond to the edges of the graph (twisted geometries framework) \cite{FreidelSpeziale:geo,FreidelSpeziale:twist}. Besides, the closure constraints are the classical counterpart of the SU(2) invariance of the intertwiners in LQG \cite{LivineTambornino}.
    
Closure constraints at one vertex commute with the spinors from the other vertex, while 
\begin{equation}
    \tinypois{\cierrex^a}{\ket{z_i}}=\frac{i}{2}\sigma^a\ket{z_i},\qquad 
    \tinypois{\cierrey^a}{\ket{w_i}}=\frac{i}{2}\sigma^a\ket{w_i}.
\end{equation}
Thus, each transformation generated by the closure constraints is a  global $\sudos$ rotation of the spinors at $\alpha $ and a global $\sudos$ rotation of the spinors at $\beta $.
Note that $\sudos$ rotations of the spinors yield $\text{SO}(3)$ rotations of the vectors defined by \eqref{vectores} and, therefore, fixing the gauge associated with these constraints implies the choice of an orientation for the set of vectors $\vec{X}_i$ on one hand, and an orientation for the set of vectors $\vec{Y}_i$ on the other.
In order to write a Hamiltonian for the model, {we define} the following $\sudos$ invariant quantities: \cite{inakiDynamics4,inakiReturn,LivineTambornino}
\begin{equation}
    E_{ij}^\alpha:=\braket{z_i}{z_j},\qquad F_{ij}^\alpha:=\dbraket{z_i}{z_j},
\end{equation}
and the analogous quantities for $\beta $.
The product of each of these quantities with its analogous at $\beta $ (for example, $E_{ij}^\alpha E_{ij}^\beta $) is also invariant under the transformations generated by the matching constraints.
Hence, the following ansatz for the Hamiltonian constraint \cite{inakiDynamics4,inakiReturn} is also invariant:
\begin{equation}\label{eq:Hamiltoniano}
    H=\sum_{i,j=1}^N \left[\lambda E_{ij}^\alpha E_{ij}^\beta+{\text{Re}}(\gamma F_{ij}^\alpha F_{ij}^\beta)\right],
\end{equation}
where $\lambda\in\real $ and $\gamma\in\cmplx$. 
	
All the constraints (matching, closure, and Hamiltonian) commute with each other.
Therefore, the phase space of the two-vertex  model (with $N$ edges) consists of $2N$ spinors reduced by $N+6+1$ first class constraints, including the Hamiltonian constraint.
Hence, the model has $8N-2\times (N+6+1)=6N-14$  real degrees of freedom.
\subsection{Twisted geometries}\label{subsec:geotuis}

The twisted geometries framework \cite{FreidelSpeziale:geo} provides a geometric parametrization for the LQG phase space on a given graph.
More concretely, the formalism gives an interpretation of this phase space in terms of polyhedra which are dual to the vertices of the graph and whose faces are dual to the edges of the graph. 
In the following, we will describe the relation between this formalism and the parametrization given by the spinorial formalism~\cite{FreidelSpeziale:twist}.
	
By virtue of Minkowski's theorem for convex polyhedra~\cite{poliedros}, given a set of $N\geq 4$ vectors $\vec{Z}_i$ that sum to zero, there is a unique convex polyhedron whose $i$-th face is normal to $\vec{Z}_i$ and has area equal to $|\vec{Z}_i|$. 
Thus, in the two-vertex model with $N$ edges, the closure constraints ensure that there is a unique association of an \mbox{$N$-face} convex polyhedron with each vertex.
Furthermore, the matching constraints impose that faces of both polyhedra associated with the same link of the graph have equal area.   
Note that, however, faces are not constrained to have the same shapes, and  hence the polyhedra do not necessarily fit together.
Nevertheless, these polyhedra do not encode all the information about the spinors that determine them.
Indeed, given the vector $\vec{X}_i$, the spinor $\ket{z_i}$ is only determined up to a global phase.
Therefore, in the twisted geometries formalism, each edge is endowed with an extra degree of freedom: the twist angle. 
	
Minkowski's theorem states that $N-1$ arbitrary vectors determine a polyhedron of $N$ faces with a fixed orientation. Thus, the space of possible polyhedra is of dimension $3(N-1)-3$, i.e., the three degrees of freedom in the $N-1$ vectors minus the degrees of freedom that encode the different orientations of the same polyhedron. 
Therefore, a pair of polyhedra with $N$ faces satisfying the matching constraints has $6(N-2)-N$ degrees of freedom and, hence, the twisted geometry of the two-vertex model with $N$ edges has $6(N-2)-N$ geometric degrees of freedom and $N$ twist angles, which yields a phase space of dimension $6(N-2)$ (with the Hamiltonian constraint yet to be imposed). 
This is in full agreement with the discussion made in Section \ref{subsec:efespin}.
\section{Two-vertex, four-edge model}\label{sec:cuatroar}
    
The two-vertex, four-edge model represents a pair of tetrahedra satisfying the matching constraints. 
Thus, it is the simplest scenario in which the polyhedra are not degenerate and we can study the dynamics of their volume and shape.
In this section we introduce a parametrization of the model with the Hamiltonian constraint yet to be imposed. We also discuss the geometric interpretation of the parameters involved and write the symplectic structure in terms of these parameters.

In what follows, we will reduce the eight spinors from the two-vertex, four-edge model by the closure and matching constraints ending with $32-2\times 10=12$ phase-space degrees of freedom.
The way in which we will fix the closure gauge is inspired by the works \cite{probing,Livineclosure}, in which only one tetrahedron is taken into account and the constraints are not completely solved in terms of the spinors. 
	
It can be deduced from \eqref{gaugedeenlace} that the gauge associated with the matching constraints can be fixed by equating the sum of the phases of the spinorial components from one and the other end of each edge:
\begin{equation}\label{gauge1}
    \arg{z_i^0}+\arg{z_i^1}-\arg{w_i^0}-\arg{w_i^1}=0.
\end{equation}
	
\begin{figure}
\tdplotsetmaincoords{60}{120} 
    \begin{tikzpicture} [scale=2.7,tdplot_main_coords, axis/.style={->,blue,thick}, 
			vector/.style={-stealth,blue,very thick}, 
			vector guide/.style={dashed,red,thick},rotate around z=20]
			
			\coordinate (O) at (0,0,0);
			

			\pgfmathsetmacro{\a}{1}
			\pgfmathsetmacro{\b}{0.5}
			\pgfmathsetmacro{\cs}{0.7757}
			\pgfmathsetmacro{\ss}{0.6319}

			\draw[vector,black] (0,0,0)  -- (0,0,-1.4) 
			node[anchor=west,pos=0.92]{$\vec{X}_3+\vec{X}_4$};
			
			\draw[vector,blue] (O) -- (\b *\cs,-\b *\ss,-0.5)node(X3)[anchor= east]{\textcolor{blue}{$\vec{X}_3$}};
			\draw[vector,blue] (O) -- (-\b *\cs,\b *\ss,-0.8)node(X4)[anchor=north west]{\textcolor{blue}{$\vec{X}_4$}};
			
			\draw[vector guide,blue]         (\b *\cs,-\b *\ss,-0.5) -- (0,0,-1.4);
			\draw[vector guide,blue]         (-\b *\cs,\b *\ss,-0.8) -- (0,0,-1.4);
			\draw[vector guide,orange] (\b *\cs,-\b *\ss,-0.5) -- (\b *\cs,-\b *\ss,0);
			\filldraw[
			draw=yellow,%
			fill=yellow!20,opacity=0.5%
			]          (1.7,1.1,0)
			-- (-1,1.1,0)
			-- (-1,-0.8,0)
			-- (1.7,-0.8,0)
			-- cycle;
			\draw[vector] (O) -- (\a *\cs,\a *\ss,0.7)node(X1)[anchor=west]{$\vec{X}_1$};
			\draw[vector] (O) -- (-\a *\cs,-\a *\ss,0.6)node(X2)[anchor=east]{$\vec{X}_2$};
			\draw[vector,orange] (0,0,0) -- (\a *\cs,\a *\ss,0)node[anchor=west]{P$\vec{X}_1$};
			\draw[vector,orange] (0,0,0) -- (\b *\cs,-\b *\ss,0)node[anchor=east]{P$\vec{X}_3$};

			\draw[vector guide,blue]         (\a *\cs,\a *\ss,0.7) -- (0,0,1.4);
			\draw[vector guide,blue]         (-\a *\cs,-\a *\ss,0.6)-- (0,0,1.4);
			\draw[vector guide,orange] (\a *\cs,\a *\ss,0.7) -- (\a *\cs,\a *\ss,0);
			\draw[axis,black] (0,0,0) -- (1.9,0,0) node[anchor=north east]{$x$};
			\draw[axis,black] (0,0,0) -- (0,1.2,0) node[anchor=north west]{$y$};
			\draw[vector,black] (0,0,0) -- (0,0,1.4) node[anchor=north, pos=1.13]{$z$} node[anchor=west,pos=0.92]{$\vec{X}_1+\vec{X}_2$};
			\tdplotdrawarc[Maroon, thick,->]{(0,0,0)}{0.6}{0}{39.13}{anchor=north}{\hspace{5pt}$ \displaystyle\frac{\varphi^\alpha}{2}$}
			\tdplotdrawarc[Maroon, thick,->]{(0,0,0)}{0.4}{0}{-39.13}{anchor=north}{\hspace{-7.5pt}$ \displaystyle\frac{\varphi^\alpha}{2}$}
	\end{tikzpicture}
	\caption{
			Closure gauge fixing at $\alpha $. The vector \mbox{$\vec{X}_1+\vec{X}_2$} is aligned to the vertical axis and pointing upwards.
			By virtue of the closure constraints, \mbox{$\vec{X}_3+\vec{X}_4$} is also aligned with the vertical axis, pointing to the opposite direction.
			It is further imposed that the projections of the vectors $\vec{X}_1$ and $\vec{X}_3$ onto the horizontal plane, P$\vec{X}_1$ and P$\vec{X}_3$, have opposite angles relative to the $x$-axis. The angle between these projections is $\varphi^\alpha$, that is, the torsion of the tetrahedron $\alpha$.
        }
	\label{fig:gauge}
\end{figure}
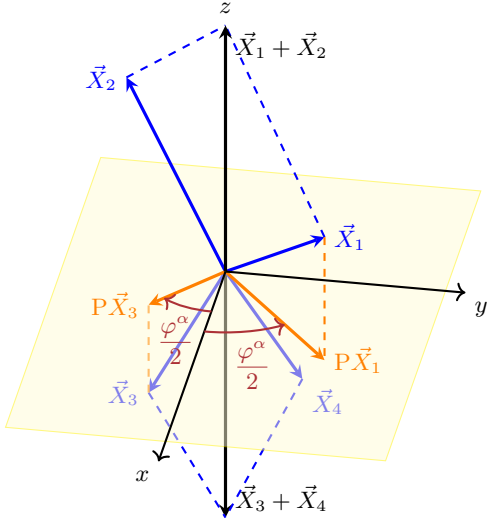

On the other hand, the gauge freedom associated with the closure constraints can be fixed by selecting an orientation for each tetrahedron.
We choose an orientation at the vertex $\alpha $ by choosing a reference frame such that the vector $\vec{X}_1+\vec{X}_2$ is parallel to the vertical $z$-axis and points upwards.
In terms of the spinors, this choice reads 
\begin{align}
    z^0_1 \overline{z}^1_1+	z^0_2 \overline{z}^1_2 & =0,\nonumber\\
    |z^0_1|^2-|z^1_1|^2+|z^0_2|^2-|z^1_2|^2&\geq 0.\label{gauge2}
\end{align} 
The closure constraints imply that the vector $\vec{X}_3+\vec{X}_4$ is parallel to the $z$-axis and points downwards (see Fig.~\ref{fig:gauge}). 
Note that the edge in which the $i$ and $j$ faces of the tetrahedron meet ($\overline{ij}$ edge for brevity) is normal to both the vectors $\vec{X}_i$ and $\vec{X}_j$ and, hence, to their sum. 
Thus, this fixation of the gauge can be understood as setting both the $\overline{12}$ and $\overline{34}$ edges of the tetrahedron to be horizontal.

Rotations around the vertical axis can be fixed by imposing the following conditions
\begin{align}
    \arg{z_1^0}-\arg{z_1^1}+\arg{z_3^0}-\arg{z_3^1}=0,\nonumber\\
    \arg{z_1^0}-\arg{z_1^1}\in\left[0,\pi\right).\label{gauge3}
\end{align} 
The former translates into the projections of $\vec{X}_1$ and $\vec{X}_3$ onto the horizontal plane having opposite angles relative to the $x$-axis (see Fig.~\ref{fig:gauge}), which amounts to the edges $\overline{12}$ and $\overline{34}$ (both horizontal) having opposite angles relative to the $y$-axis. 
The latter imposes the second component of the vector $\vec{X}_1$ to be positive.
	
We proceed in an analogous manner at the vertex $\beta $ so that we completely fix the gauge.
Note that the conditions  \eqref{gauge1}-\eqref{gauge3} (and the analogous equations at $\beta $) sum up to ten real equations.
    
Once the gauge has been fixed in this way, the spinors which satisfy the matching and closure constraints can be written in terms of twelve parameters $A_i$, $x^{\alpha,\beta}$, $\Phi_i$, and $\varphi^{\alpha,\beta}$ ($i=1,2,3,4$).
Concretely, the spinors at the vertex $\alpha $ can be parametrized as
\begin{align}
    &\ket{z_1} = e^{{i\Phi_1/4}}
    \begin{pmatrix}
        f^{(12)\alpha}_+ \\
        f^{(21)\alpha}_- 
        \end{pmatrix},
    &&\ket{z_3} = e^{{i\Phi_3/4}}
    \begin{pmatrix}
        f^{(43)\alpha}_-  \\
        f^{(34)\alpha}_+ 	
    \end{pmatrix},\nonumber\\
    &\ket{z_2} = e^{{i\Phi_2/4}}
    \begin{pmatrix}
        f^{(21)\alpha}_+ \\
        -f^{(12)\alpha}_- 
    \end{pmatrix},
    &&\ket{z_4} = e^{{i\Phi_4/4}}
    \begin{pmatrix}
        f^{(34)\alpha}_-  \\
        -f^{(43)\alpha}_+	
    \end{pmatrix},\label{parafinal}
\end{align}
with
\begin{equation}
    f^{(ij)\alpha}_\pm :=
    \sqrt{
        \frac{
            \left(A_{ij}\pm x^\alpha \right)
		  \left(x^\alpha+D_{ij}\right)
        }
	   {2x^\alpha }
    }e^{{\pm i \varphi^\alpha/4}},
\end{equation}
where
\begin{equation}
    A_{ij}:=A_i+A_j,\qquad
    D_{ij}:=A_i-A_j.
\end{equation} 
The spinors at $\beta $ are written analogously, using $x^\beta $ and $\varphi^\beta$ instead of $x^\alpha $ and $\varphi^\alpha$.
    
The quantities $A_i$ and $x^{\alpha,\beta}$ are real, positive, and constrained by the following inequalities:
\begin{equation}\label{eq:desigualdad}
    \max\left(|D_{12}|,|D_{34}|\right)\leq x^{\alpha,\beta}\leq
    \min\left(A_{12},A_{34}\right),
\end{equation}
while the remaining parameters, $\Phi_i, \varphi^{\alpha,\beta}\in\left[0,2\pi\right)$, are angles.
All these variables can be identified as gauge invariant quantities from the model, as we detail in what follows.
\begin{itemize}[leftmargin=0.3cm]
    \item $A_i$ are the areas associated with the edges of the graph $A_i=\frac{1}{2}E^\alpha_{ii}=\frac{1}{2}E^\beta_{ii}$.
    \item $x^{\alpha}=|\vec{X}_1+\vec{X}_2|$ is the area of the shadow that the tetrahedron $\alpha $ would cast on the horizontal plane.  
        In virtue of the closure constraint, we can also write $x^{\alpha}=|\vec{X}_3+\vec{X}_4|$. 
	The analogous situation is held at~$\beta $.
	The inequalities \eqref{eq:desigualdad} follow from this definition of~$x^{\alpha,\beta}$.
	These quantities can be related to the interior dihedral angle between the faces $1$ and $2$ at $\alpha$ or $\beta $ as follows:
	\begin{equation}\label{eq:diedrico}
	   \cos(\theta_{12}^{\alpha,\beta})=\frac{(A_1)^2+(A_2)^2-(x^{\alpha,\beta})^2}{2A_1A_2}.
	\end{equation}
	The analogous relation to the dihedral angle between the faces $3$ and $4$ is satisfied as well.
	With all the other variables fixed, the tetrahedra elongate along the vertical axis as $x^{\alpha,\beta}$ have smaller values, and the other way around. 
	Hence, we shall call these quantities the \emph{flattenings} of the tetrahedra. 
    \item The angles $\Phi_i$ can be directly related to the quantities with which the Hamiltonian is constructed:
        \begin{align}
            \Phi_1&=\arg(E^\alpha_{21}E^\beta_{21}F^\alpha_{12}F^\beta_{12}),\nonumber\\
            \Phi_2&=\arg(E^\alpha_{12}E^\beta_{12}F^\alpha_{12}F^\beta_{12}),\nonumber\\
            \Phi_3&=\arg(E^\alpha_{43}E^\beta_{43}F^\alpha_{34}F^\beta_{34}),\nonumber\\
            \Phi_4&=\arg(E^\alpha_{34}E^\beta_{34}F^\alpha_{34}F^\beta_{34}).
	\end{align}
	These angles parametrize all the information from the model which cannot be read from the tetrahedra. Hence, we will identify them as the twist angles.
	In fact, in this gauge, they are proportional to the sum of the two valid definitions of the twist angle given in~\cite{FreidelSpeziale:twist}: 
	\begin{equation}
		\Phi_i=\arg{z_i^0}+\arg{w_i^0}+\arg{z_i^1}+\arg{w_i^1}.
	\end{equation}
    \item Finally, $\varphi^\alpha$ is the angle between the projections of $\vec{X}_1$ and $\vec{X}_3$ onto the horizontal plane (normal to $\vec{X}_1+\vec{X}_2$, see Fig.~\ref{fig:gauge}) and, therefore, can be defined in terms of scalar products of the vectors at $\alpha$, which are gauge invariant.
        Geometrically, the angle $\varphi^{\alpha}$ measures the rotation of the $\overline{12}$ edge relative to the $\overline{34}$ edge.
        For this reason, we will refer to $\varphi^{\alpha}$ as the torsion of the tetrahedron at the vertex $\alpha$. 
        $\varphi^\beta$ is the analogous quantity in the vertex $\beta $.
\end{itemize}

Furthermore, this is a canonical parametrization:
\begin{equation}
    \pois{A_i}{\Phi_j}=\delta_{ij},\ \pois{x^\alpha}{\varphi^\alpha}=\pois{x^\beta}{\varphi^\beta}=1,
\end{equation}
and any other bracket between variables is zero. 
This can be checked writing the symplectic term of the action, $i\sum_{i=1}^4(\overline{z}^A\dot{z}^A+\overline{w}^A\dot{w}^A)$, in the presented parametrization. 
The areas and twist angles being canonical conjugates corresponds to the symplectic structure with which the twisted geometries are equipped \cite{FreidelSpeziale:geo}. 
On the other hand, we can write $x^\alpha = \frac{1}{2}(\vec{X}_1+\vec{X}_2-\vec{X}_3-\vec{X}_4)\cdot \vec{u}_{12}$, where $\vec{u}_{12}$ is the unit vector in the (vertical) $\vec{X}_1+\vec{X}_2$ direction, and check that, via Poisson brackets, $x^\alpha $ generates rotations around $\vec{u}_{12}$ in one direction for the vectors 1 and 2 and in the opposite direction for the vectors 3 and 4 wringing the tetrahedron. 
Hence, it is natural that the flattening and torsion of each tetrahedron are canonical conjugates.

Additionally, let us note that, following~\cite{probing}, Eq.~\eqref{eq:diedrico} yields
\begin{equation}\label{eq:volumen}
    V_\alpha ^2=\frac{2A_1A_2A_3A_4}{9x^\alpha}|\sin(\varphi^\alpha)\sin(\theta_{12}^\alpha)\sin(\theta_{34}^\alpha)|
\end{equation}
for the volume of the tetrahedron $\alpha$ in terms of our canonical parametrization.
Hence, it is possible to study the dynamics of the volume of the tetrahedra within the presented parametrization. 
The expansion of the volume can be related to the mean expansion of each tetrahedron. Its relation to the expansion that each area undergoes determines whether the tetrahedron is deformed by the dynamics of the system. 
\section{Homogeneous and isotropic sector}\label{subsec:uene} 

In previous works a reduced sector within the two-vertex model was studied: the so-called $\uene $ sector, which is naturally identified as the homogeneous and isotropic sector within the graph \cite{inakiReturn,inakiDynamics4,BorjaGarayVidotto_learning,alvaro2024,BenitoLivine:classicalsetting}. 
Furthermore, it turns out that the dynamics of the model in this sector reproduce the old LQC dynamics \cite{alvaro2024}. 
In what follows we will study how this reduced sector is characterized in terms of the parametrization \eqref{parafinal}.

The $\uene$ sector is defined as the reduction by the constraints \cite{inakiReturn}
\begin{equation}\label{eq:liguene}
    \mathcal{E}_{ij}:=E_{ij}^\alpha-E_{ji}^\beta=0,
\end{equation}
which commute with the Hamiltonian and with the closure constraints in general, and weakly commute with the matching constraints:
\begin{equation}
    \pois{\mathcal{E}_{ij}}{\mathcal{C}_k}=i(\delta_{ik}-\delta_{jk})\mathcal{E}_{ij}.
\end{equation}
Therefore, the $\uene $ constraints are both stable under the evolution of the system and compatible with any fixation of the matching and closure gauges.

Note that these constraints can be written as
\begin{equation} 
    \braket{z_i}{z_j}=\braket{\overline{w}_i}{\overline{w}_j}. 
\end{equation}
Therefore, in the $\uene$ sector there is a unitary transformation $u\in \text{U}(2)$ such that $\forall i,\ u\ket{\overline{w}_i}=\ket{z_i}$. 
We can write $u=e^{i\Phi/2}h$, with $h\in \sudos$.
In the parametrization \eqref{parafinal}, $h$ can only be a rotation around the vertical axis, since the spinors at both vertices satisfy the conditions \eqref{gauge2}, which are only preserved by rotations of said kind.

Hence, the $\uene$ sector, in terms of our parametrization, is the one in which there exists a pair of angles $\Phi$ and $\varphi $ such that $\forall j,\ e^{i\Phi/2}e^{i(\varphi/2)\sigma^3}\ket{\overline{w}_j}=\ket{z_j}$, where $\sigma^3$ is the usual third Pauli matrix.
    
Therefore, it is straightforward to check that the $\uene$ reduction is equivalent to imposing the following five first-class constraints:
\begin{align}\label{eq:homogeneidad}
    x^\alpha-x^\beta&= 0,\\
    \label{eq:isotropia1}
    \varphi^\alpha+\varphi^\beta&= 0,\\\label{eq:isotropia2}
    \Phi_1-\Phi_2= 
    \Phi_2- \Phi_3&= 
    \Phi_3-\Phi_4=0.
\end{align}
These constraints allow the existence of the aforementioned unitary transformation, with 
\begin{equation}
    \Phi=\frac{1}{4}\sum_{i}\Phi_i,\qquad 
    \varphi=(\varphi^\alpha-\varphi^\beta)/2.
\end{equation}
From the symplectic structure it follows that $\varphi^\alpha-\varphi^\beta$, $x^\alpha+x^\beta $, and the difference between the areas of any two edges are pure gauge in the $\uene$ sector. 
All the relevant information is contained in the canonically conjugate pair made of the total area $A:=\sum_{i}A_i$ and the average twist angle, $\Phi$, as it is noted in \cite{inakiDynamics4,alvaro2024,BenitoLivine:classicalsetting}. 
These variables are global: they do not distinguish one vertex from the other or any edge from the rest, which is why it is stated that the $\uene $ sector corresponds to the intuitive notion of homogeneity and isotropy in the graph.
    
Note that, once the constraint~\eqref{eq:homogeneidad} is imposed, the tetrahedra have the same flattenings and, modulo gauge transformations, the same torsions. Therefore, this constraint makes any distinction between the vertices irrelevant, being responsible for the homogeneity of the sector.  
   
Let us analyze the meaning of the remaining four constraints. By virtue of the constraint~\eqref{eq:isotropia1}, the mean torsion of the tetrahedra is fixed to zero, while the mean flattening becomes irrelevant. This removes two degrees of freedom of the shape that the tetrahedra (modulo gauge transformations) have in common. 
The result of the imposition of the constraints~\eqref{eq:homogeneidad} and~\eqref{eq:isotropia1} is that the tetrahedra are mirror images of each other.
The constraints~\eqref{eq:isotropia2} make all the twist angles equal and any difference between areas irrelevant, eliminating any distinction between the edges of the graph. Moreover, they cause the remaining freedom of the shape of the tetrahedra (associated with the independence between face areas) to be gauge. Thus, the constraints~\eqref{eq:isotropia1} and~\eqref{eq:isotropia2} are responsible for the isotropy. 
    
From this analysis one can check, imposing the constraints in the parametrization \eqref{parafinal}, that the Hamiltonian in this sector is
\begin{equation}
    H_\text{red}=2A^2(\lambda+\gamma\cos\Phi),
\end{equation}
which agrees with the results in \cite{inakiDynamics4,inakiReturn}. 
Note that we have assumed $\gamma\in\real^+$, since the complex phase of this constant can be absorbed as a redefinition of $\Phi $. 
In fact, the phase of $\gamma $ can always be absorbed by redefining the twist angles $\Phi_i $. Hence, we will consider \mbox{$\gamma\in\real^+$} from now on.
\section{Anisotropic and inhomogeneous sectors}\label{sec:sectores}
    
The analogy between the homogeneous and isotropic sector of the two-vertex model and LQC suggests the possibility of using this family of simple graphs to describe less symmetric cosmological scenarios.  
Hence, it is interesting to study less restrictive reduced sectors with more degrees of freedom than the $\uene$ symmetric model in order to study whether the analogy extends to inhomogeneous or anisotropic cases.
	
In this section we will present some reduced sectors (with more degrees of freedom than the $\uene$ sector) which are stable under evolution. We will use the parametrization given in Section~\ref{sec:cuatroar} to identify the new degrees of freedom as degrees of anisotropy or inhomogeneity in the same fashion that the $\uene$ sector is identified as the homogeneous and isotropic sector of the model.
\subsection{Sector with a privileged direction}\label{udp}
    
As we have discussed, one of the reasons for the $\uene$ sector to be considered isotropic is that none of the edges of the graph evolves independently. 
Hence, a natural way to find a reduced sector with some degree of anisotropy is to decouple one of the edges of the graph from the definition of the $\uene $ sector. 
In other words, we can study the sector which is reduced only by the constraints $\mathcal{E}_{ij}$ with $i\neq 1$ and $j\neq 1$, decoupling the edge $1$ and, hence, establishing a privileged direction.
Each $\uene $ constraint commutes with the Hamiltonian independently, therefore a sector reduced only by some of them, as the case we are considering, is stable under evolution. 
    
Analogously to the $\uene $ case, this is equivalent to demand that there exists some transformation with the form $e^{i\Phi/2}e^{i(\varphi/2){\sigma^3}}$ mapping the spinors $\ket{\overline{w}_2},\ \ket{\overline{w}_3},\ \ket{\overline{w}_4}$ to the spinors $\ket{z_2},\ \ket{z_3},\ \ket{z_4}$.
It follows directly that this reduction can be achieved by imposing the following four first-class constraints:
\begin{align}
    x^\alpha-x^\beta&=0,\\ 
    \varphi^\alpha+\varphi^\beta&= 0,\\ 
    \Phi_2-\Phi_3= 
    \Phi_3-\Phi_4&=0.
\end{align}
Thus, as happened in the $\uene$ sector, the tetrahedra are mirror images of each other, but now only the twist angles 2, 3, and 4 are made equal, while $\Phi_1$ remains independent. 
Therefore, following the interpretation of the constraints explained in the previous section, we have kept homogeneity while breaking isotropy.
	
All the relevant information in this sector can be parametrized with the canonical pairs $(A_1,\Phi_1)$ and $(A_I,\Phi_I)$, where $A_I:=\sum_{i=2}^4A_i,\ \Phi_I:=\frac{1}{3}\sum_{i=2}^4\Phi_i$. 
Contrary to what we had in the $\uene$ sector, there is some relevant freedom in the shape of the tetrahedra: if two configurations have different values for $A_1$ two different points of the phase space correspond to them, even if the total area $A$ is the same in both configurations.  
    
When writing the Hamiltonian in this sector it is convenient to use the following canonical pairs $(A,\Phi_{\text{u}})$ and $(D_{\text{u}},\Delta_{\text{u}})$ instead, where $A$ is the total area and
\begin{equation}
    \Phi_{\text{u}}:=\frac{1}{2}\left(\Phi_I+\Phi_1\right),\ D_{\text{u}}:=A_I-A_1,\ \Delta_{\text{u}}:=\frac{1}{2}\left(\Phi_I-\Phi_1\right).
\end{equation}
The Hamiltonian defined in \eqref{eq:Hamiltoniano} has, then, the following expression in this sector:
\begin{equation}
    \begin{aligned}
        H_{\text{u}}&=2A^2\left(\lambda +\gamma \cos{\Phi_{\text{u}}}\right)
        -2\gamma AD_{\text{u}}\sin{\Phi_{\text{u}}}\sin{\Delta_{\text{u}}}\\
        &-2D_{\text{u}}\left[\lambda (A-D_{\text{u}})+\gamma A\cos{\Phi_{\text{u}}}\right](1-\cos{\Delta_{\text{u}}}).
    \end{aligned}
\end{equation}
Note that $\Phi_{\text{u}}$ coincides with $\Phi $ in the $\uene$ sector, which is recovered when $\Delta_{\text{u}}=0$, and that the first term that appears in this expression is the $\uene $ Hamiltonian (substituting $\Phi$ by $\Phi_\text{u}$). 
Thus, we can read this Hamiltonian as the homogeneous and isotropic Hamiltonian plus some correction due to the new anisotropic degree of freedom.
This correction depends on the global parameters $A$ and $\Phi_\text{u}$, hence the degrees of anisotropy $D_\text{u}$ and $\Delta_\text{u}$ modify the evolution of the $\uene $ variables. 
    
This sector constitutes a simple scenario in which the cosmological interpretation of the two-vertex model can be further studied, since it only adds a pair of degrees of freedom to the  $\uene$ sector.
\subsection{Bi-twist sector}
In this section we will impose the constraints $\mathcal{E}_{12}$ and $\mathcal{E}_{34}$ only. 
Analogously to the previous cases, this leads to a stable sector, in which there must be angles $\Phi_{12},\ \Phi_{34},\ \varphi_{12}$ and $\varphi_{34}$ such that
\begin{equation}
    e^{i{\Phi_{12}/2}}e^{i{(\varphi_{12}/2)}{\sigma^3}}\ket{\overline{w}_{1,2}}=\ket{z_{1,2}},
\end{equation}
and equivalently for the edges $3$ and $4$. 
Hence $\mathcal{E}_{12}=~\mathcal{E}_{34}=0$ is equivalent to the following three first-class constraints:
\begin{align}
    x^\alpha-x^\beta&=0,\label{bituis}\\
    \Phi_1-\Phi_2= \Phi_3-\Phi_4&=0.\label{bituis2}
\end{align}
Once again, homogeneity is maintained (Eq. \eqref{bituis}), while the twist angles have been equated in pairs instead, imposing ``isotropy'' only between the edges 1 and 2 on the one hand, and between the edges 3 and 4 on the other.
Thus, in this sector two twist angles and two areas contain the relevant information related to the edges of the graph.
Note that Eq.~\eqref{bituis} implies that the difference between torsions $\varphi^\alpha-\varphi^\beta$ is pure gauge and no further constraints are imposed on them. Furthermore, Eq.~\eqref{bituis2} implies that the area differences $A_1-A_2$ and $A_3-A_4$ are also pure gauge.
    
Therefore, this sector differs from the previous one in that an additional pair of degrees of freedom related to the shape of the tetrahedra arises (this pair being the average flattening and average torsion). In other words, this sector has one more degree of anisotropy.
    
We will call this sector the bi-twist sector and it can be parametrized by the canonical pairs $(A,\Phi )$, $(D,\Delta)$, $(x,\varphi)$, where $A$ is the total area, $\Phi$ the average {twist} angle, and
\begin{align}
    x&:=x^\alpha+x^\beta,\qquad \varphi:=\frac{1}{2}\left(\varphi^\alpha+\varphi^\beta\right),\nonumber\\
    D&:=A_1+A_2-A_3-A_4,\nonumber\\
    \Delta&:=\frac{1}{4}\left(\Phi_1+\Phi_2-\Phi_3-\Phi_4\right).
\end{align}
The Hamiltonian in this sector acquires the form
\begin{align}
    H_\text{b}&=2A^2(\lambda+\gamma\cos\Phi)-
    \lambda B_-(1-\cos\Delta\cos\varphi)\nonumber\\
    &-\gamma\cos\Phi\left[B_+(1-\cos\Delta)+B_-(1-\cos\varphi)\right]\nonumber\\
    &+2\sin\Phi\left[\lambda Dx\sin\Delta+\gamma A( D\sin\Delta+x\sin\varphi)\right],
\end{align}
where $B_\pm=A^2\pm D^2-x^2$. The first term is the $\uene $ Hamiltonian, while the remaining terms disappear in the $\uene$-limit, i.e., when imposing $\Delta=\varphi=0$.
Once again, we can identify the new terms with the additional  anisotropic degrees of freedom that appear in this sector.
    
Note that the bi-twist sector is not contained in the one privileged direction sector, and vice versa. In fact, their intersection is the $\uene$ sector.
    
The dynamics of the bi-twist sector was numerically studied in \cite{ArangurenGarayLivine}, although it was done without a priori knowledge about the stability of the sector, which was proposed as an ad hoc parametrization for the case in which the areas are equated in pairs. 
Indeed, the parametrization presented in the section III of \cite{ArangurenGarayLivine} corresponds to the imposition of the constraints \eqref{bituis} and the gauge fixing $A_1-A_2=A_3-A_4=0$ (and a different fixation of the gauges of matching and closure, in which the gauge freedom of global rotations around the vertical axis remains).
\subsection{Inhomogeneous bi-twist sector } 
    
We will now present a sector arising from not imposing the homogeneity condition \eqref{bituis} in the bi-twist sector: the inhomogeneous bi-twist sector.
Concretely, we will consider the following set of second class constraints:
\begin{equation}
	A_1-A_2=A_3-A_4=\Phi_1-\Phi_2=\Phi_3-\Phi_4=0.
\end{equation}
It turns out that these constraints weakly commute with the Hamiltonian of the model. Hence, they define a stable sector, in which the areas and twist angles are equated in pairs, while no condition is imposed on the flattenings and torsions, leaving room for the two tetrahedra to differ from one another.
    
The inhomogeneous bi-twist sector can be parametrized in terms of the canonical pairs $(x,\varphi)$, $(y,\delta)$, $(A,\Phi )$, $(D,\Delta)$, where, additionally to the previous definitions, we have
\begin{equation}
	y:=x^\alpha-x^\beta,\qquad \delta:=\frac{1}{2}(\varphi^\alpha-\varphi^\beta).
\end{equation}
Note that this pair carries information about the inhomogeneity of each configuration. 
The Hamiltonian takes the following form:
\begin{align}
    H_\text{bi}&=\lambda\left[A^2+2D^2+2x^2-2y^2\right.\nonumber\\
    &+\!\left.g(D\!+\!X,A)\cos(\Delta\!+\!\varphi)
    +
    g(D\!-\!X,A)\cos(\Delta\!-\!\varphi)\right]
    \nonumber\\
    &+\!\gamma\left[g(A\!+\!X,D)\cos(\Phi\!+\!\varphi)+
    g(A\!-\!X,D)\cos(\Phi\!-\!\varphi)\right.
    \nonumber\\
    &\left.
    +g(A\!+\!D,X)\cos(\Phi\!+\!\Delta)
    +g(A\!-\!D,X)\cos(\Phi\!-\!\Delta)
    \right],
\end{align}
where
\begin{equation}
	g(a,b):=\frac{1}{2}
    \sqrt{
	   \left(a^2-(b+y)^2
	   \right)
	   \left(a^2-(b-y)^2
	   \right)
	}.
\end{equation}
    
Note that the Hamiltonian does not depend on $\delta $. Thus $y$ is a conserved quantity. 
Whenever this quantity has a non-vanishing value, the dynamics is notably different from those in the previous sectors, which all fulfill the condition $y=0$. Thus, this sector provides an arena where we may explore the cosmological interpretation of the model for inhomogeneous scenarios.
    
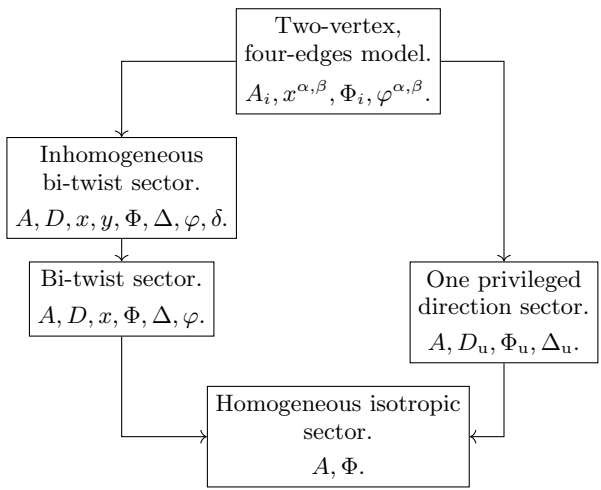
\begin{figure}[ht]
    \begin{tikzpicture}[node distance=0.25cm]
					
        \matrix[matrix of nodes,
            column sep =-0.4cm, row sep =0.3cm,every node/.style={anchor=north}] 
            {\node(a){ };&\node(general)[draw,align=center]{Two-vertex,\\four-edges model.\vspace{3pt}\\ $A_i,x^{\alpha,\beta},\Phi_i,\varphi^{\alpha,\beta}.$};\\
		  \node(bih)[draw,align=center]{Inhomogeneous\\bi-twist sector.\vspace{3pt}\\ $A,D,x,y,\Phi,\Delta,\varphi,\delta.$};\\
		  \node(bi)[draw,align=center]{Bi-twist sector.\\\vspace{-7pt} \\$A,D,x,\Phi,\Delta,\varphi.$};
		  & &&\node(udp)[draw,align=center]{One privileged\\direction sector.\vspace{3pt}\\ $A,D_\text{u},\Phi_\text{u},\Delta_\text{u}.$};\\
		  & \node(uene)[draw,align=center]{Homogeneous isotropic\\sector.\vspace{3pt}\\ $A,\Phi.$};\\};     
		\path[draw,->](general) -| (bih);
		\path[draw,->](general)-|(udp);
		\path[draw,->](bih)--(bi);
		\path[draw,->](bi)|-(uene);
		\path[draw,->](udp)|-(uene);
	\end{tikzpicture}
	\caption{
        Relations between the sectors presented in Section~\ref{sec:sectores}.
		Each arrow represents the imposition of some constraints, so that the result is a subsector of the sector from which starts.
		Below the name of each sector, we display the degrees of freedom  that characterize it.
    }
	\label{jerarquia}
\end{figure}

In Fig.~\ref{jerarquia} we diagrammatically present the relations between the different stable sectors that have been found throughout this work.

\section{Conclusions}\label{sec:Conclusion}

In previous works~\cite{inakiDynamics4,BorjaGarayVidotto_learning,inakiReturn,BenitoLivine:classicalsetting,ArangurenGarayLivine,alvaro2024}, the LQG reduced models associated with graphs consisting of two vertices linked by $N$ edges were studied. In these works, the framework given by the twisted geometries and the spinorial formalism  for LQG ~\cite{LivineTambornino,FreidelSpeziale:geo} played an important role to propose suitable dynamics and explore the evolution of the model.	A $\uene$-symmetry reduced sector which is stable under evolution was identified. It was also shown that this sector can be interpreted as its homogeneous and isotropic sector both kinematically and dynamically~\cite{inakiReturn,BenitoLivine:classicalsetting}, which indicates that  it may successfully describe FLRW cosmology (even with effective LQC corrections)~\cite{alvaro2024}.
    
In this work, we have focused on the two-vertex model with four edges, aiming to characterize its behavior in general, outside the $\uene $ sector. 
According to the twisted-geometries formalism, it represents a pair of tetrahedra, being the simplest non-trivial example of the family of two-vertex graphs. 
Starting from the spinorial formalism, we have found a canonical parametrization of the model in which the closure and matching constraints are completely solved, meaning that there only remain the relevant degrees of freedom. 
We have also fully identified these degrees of freedom: eight of them are geometric quantities that determine the pair of tetrahedra, and the remaining four are the twist angles.
Furthermore, this parametrization enables a precise understanding of how homogeneity and isotropy arise from the reduction to the $\uene$ sector, which is key to understanding the role of the degrees of freedom found on more general scenarios. 
 	
Finally, we have found three symmetry-reduced sectors which are stable under evolution and have the $\uene$ sector as a subsector: the sector with one privileged direction, the bi-twist sector, and the inhomogeneous bi-twist sector.
Moreover, the additional degrees of freedom (compared with the $\uene$ reduction) of each sector can be interpreted as anisotropies or inhomogeneities in the same sense that the $\uene$ sector is understood to be homogeneous and isotropic. This will be very useful as guidance when comparing the dynamics of these new sectors to those of known anisotropic or inhomogeneous cosmological models.
\acknowledgments

We are grateful to Álvaro Cendal, Sergio Rodríguez, and Raül Vera for enlightening conversations.
    
Financial support was provided by the Spanish Government Grants No. PID2023-149018NB-C44 and No. PID2021-123226NB-I00 (funded by
MCIN/AEI/10.13039/501100011033  and by ``ERDF A way of making Europe''), the Basque Government
Grant No. IT1628-22, and the Natural Sciences and Engineering Research Council of Canada (NSERC). 
\bibliography{biblio}
\end{document}